\documentstyle[12pt]{article}
\begin{document}
\baselineskip 7mm
\title{ Fractal Wilson Loop - Area Law and Gauge Invariance in Next to Leading
Order }
\author{ D. Allouani$^{2}$ and H. Kr\"oger$^{1,2}$ \\ [2mm]
{\small\sl $^{1}$ Max Planck Institut f\"ur Kernphysik,
P.O.B. 10 39 80, D-6900 Heidelberg, Germany } \\
{\small\sl $^{2}$ D\'epartement de Physique, Universit\'e Laval,
Qu\'ebec, Qu\'ebec G1K 7P4, Canada }}
\date{ April 1994 }
\maketitle
\begin{flushleft}
{\bf Abstract}
\end{flushleft}
We consider a fractal Wilson loop $<F_{P}>$ and present physical arguments
why this should be a relevant observable in nature.
We show for non-compact SU(2) lattice gauge theory in the next to leading order
of strong coupling expansion that $<F_{P}>$ obeys
an area law behavior and is gauge invariant.
\begin{flushleft}
PACS index: 11.15.Ha
\end{flushleft}
\setcounter{page}{0}
\maketitle

\newpage
\section{ Motivation }
Wilson \cite{kn:Wils74} suggested in 1974
the lattice regularisation of gauge theory in terms of compact link variables.
Wilson's main argument was, in order to provide confinement, that the theory
should be manifestly gauge invariant already on the classical level.
Wilson introduced an order parameter to distinguish the confined phase, the
Wilson loop $<W_{P}>$. It corresponds to the amplitude to create a
quark-antiquark pair, propagate it and annihilate it somewhere else.
In order to relate the Wilson loop
to the notion of a quark-antiquark potential, one has to make the assumption
that the quarks propagate classically,
i.e., they are heavy and they move along smooth (differentiable) curves
\cite{kn:Band81,kn:Kogu83}.

\bigskip

One may ask the question: What happens in the case of light quarks,
e.g., $q_{u}$, $q_{d}$, which are also confined according to experiment?
In order to decide if a quark is light, one can compare
it with the dynamical scale $\Lambda$ of QCD.
Gribov \cite{kn:Grib87} has proposed a confinement mechanism, which is entirely
based upon the presence of light quarks.
Let us hence consider light quarks. Then there is no reason, why they should
propagate on smooth classical trajectories. It is more realistic to assume that
its dynamics is governed by quantum fluctuations.
What can be said more specifically on the nature of those quantum fluctuations
and on the geometry of trajectories?

\bigskip

Let us discuss this using the language of path integrals.
{}From Feynman and Hibbs \cite{kn:Feyn65}
it is known that in quantum mechanics individual pathes can be viewed
as zig-zag lines, which are no-where differentiable.
These pathes are fractal curves of Hausdorff dimension two
\cite{kn:Abbo81}. In imaginary time (Euclidean) quantum mechanics the
free motion resembles very much the Brownian motion of a classical particle
\cite{kn:Nels66}. It turns out that the average path
of a classical particle carrying out a Brownian motion (steps $\Delta x$
and $\Delta t$, such that the diffusion coefficient
$D=\frac{1}{2} \Delta x^{2} / \Delta t$ is finite)
is also a fractal curve of Hausdorff dimension two.
The Euclidean path integral of imaginary time quantum mechanics
corresponds to a Wiener measure, as discussed by Schulman \cite{kn:Schu81}.
The case of the free scalar field $\phi$ is discussed by Itzykson and Drouffe
\cite{kn:Itzy89}.
Quite generally, the measure of Euclidean path integrals
gives the dominant contributions coming from field configurations
corresponding to pathes of no-where differentiable curves \cite{kn:Glim81}.

\bigskip

Now let us return to the propagator of light quarks.
In the view of the above remarks on quantum mechanical pathes,
it seems plausible to assume
that there will be important contributions due to zig-zag curves.
Why do we care about the zig-zag-ness of paths?
The answer is: Usually, when computing the quantum expectation value of an
observable one simply sums over the paths.
However, there may be non-local observables, which are sensitive to the
zig-zag-ness.
Based on this the concept of a fractal Wilson loop
$<F_{P}>$ has been proposed \cite{kn:Krog92,kn:Krog93}.
It might be interesting to do experiments on the foundations of quantum
mechanics, e.g., in atomic physics,
looking for zig-zag-ness via a suitable observable. To the author's knowledge,
no such experiment has been done yet.

\bigskip

In order to understand the reason for a fractal Wilson loop, let us
recall the construction of the standard Wilson loop.
In order to get a quark-antiquark potential interpretation,
one starts to construct the Wilson loop from a meson-meson
correlator
$< 0 \mid M(x_{1},x_{2}) M^{\dagger} (y_{1},y_{2}) \mid 0 >$,
where the creation operator with meson quantum numbers
is built from a quark-antiquark pair,
$M(x,y) = q(x) \exp[i g \int_{x}^{y} dz A(z) ] \bar{q}(y)$.
The standard Wilson loop is obtained by integrating out the fermions,
by making the assumption that they behave classically.
Thus one ends up with a gluonic path integral, and the observable being
$\exp[ i \; g \; dz_{\mu} A_{\mu}(z) ]$ along a smooth closed curve.
But from the above point of view taking serious the quantum mechanics
via path integrals  means to sum over fermionic zig-zag lines.
Thus we are led to an observable constructed along a fractal curve,
i.e., the fractal Wilson loop.

\bigskip

Why do we expect that a fractal Wilson loop might eventually show
a behavior different from a standard Wilson loop?
A linear confining potential is equivalent to an area law behavior
of the Wilson loop.
On the other hand, a fractal curve of Hausdorff dimension $d_{H}=2$
behaves in a way as a two dimensional surface.
Thus a fractal Wilson loop defined along a curve of Hausdorff dimension
two might display an area law behavior.

\bigskip

If the concept of a fractal Wilson loop is to taken serious,
it has to fulfil two criteria:
(a) area law behavior, (b) gauge invariance.
In Ref. \cite{kn:Krog92} it has been shown to lowest order of the strong
coupling expansion that both criteria are fulfilled for $SU(2)$ non-compact
lattice action.
More precisely, although the classical action and the classical fractal Wilson
loop $F_{P}$ are not gauge invariant, the quantum mechanical expectation
value $<F_{P}>$ is gauge invariant.
One should note that numerical lattice simulations
of the standard Wilson loop, using the non-compact $SU(2)$ action
have not given an area law, but a perimeter law behavior as has been observed
by several workers \cite{kn:Patr81}.

\bigskip

As pointed out, for non-compact $SU(2)$ action there is
a distinction between standard and fractal Wilson loop, at least for
lowest order of strong coupling. Now we want to make one step further and
present new calculations extending the result of
an area law behavior and gauge invariance of the fractal Wilson loop
to the next to leading order.

\section{ Fractal Wilson loop $<F_{P}>$ }
In order to get a mathematically well defined geometry,
we choose a construction employing self-similarity,
in a way such that in the classical continuum limit $a \rightarrow 0$
the fractal Wilson loop coincides with the standard Wilson loop.
The construction of $F_{P}$ is given in Ref.\cite{kn:Krog92}.
We present the basic notation.
We denote the link between two neigboring sites $ij$ and a plaquette between
four neighboring sites $ijkl$ by
\begin{eqnarray}
U_{ij} & = & \exp [ i g \Delta x_{\mu} A^{a}_{\mu}(center \; ij) \tau^{a} ],
\nonumber \\
V_{ijkl} & =& \exp [ i g \Delta \sigma_{\mu \nu} F^{a}_{\mu \nu}
(center \; ijkl) \tau^{a} ].
\end{eqnarray}
One has the relation
\begin{equation}
V_{ijkl} = U_{ij} U_{jk} U_{kl} U_{li} + O((\Delta x)^{4}).
\end{equation}
The fractal Wilson loop $F_{P}$ is obtained from a standard Wilson loop $W_{P}$
by growing 'branches of links' into the interior, and whenever possible,
converting four neighboring sequential links $U$ into a plaquette $V$.
Thus one arrives at well defined ordered product of links and plaquettes
\begin{equation}
F_{P} = Tr \left[ p.o. + s.o. \prod_{\partial P_{fract}} \prod_{P_{fract}}
\{ U, V \} \right],
\end{equation}
where $P_{fract}$ is the surface covered by the elementary plaquettes
$(\Box)$ and $\partial P_{fract}$ is the interior boundary covered by links
$(-)$ (see Fig.[1]).
Simple geometrical consideratons show, when $a/L \rightarrow 0$,
that the area of $P_{fract}$ obeys
$A_{P_{fract}} / A_{P} \rightarrow 2/3$, and that $\partial P_{fract}$ is a
fractal curve with fractal dimension $D_{fract}=2$.

\section {Area law behavior}
We consider the functional integral over the gauge field,
using rescaled field variables $\vec{B}_{\mu} = \sqrt{g} a \vec{A}_{\mu}$,
$\vec{f}_{\mu \nu} = a^{2} \vec{F}_{\mu \nu}$. We define
$\epsilon =1/\sqrt{g}$, thus
\begin{eqnarray}
\vec{f}_{\mu \nu} &=& \vec{f}^{(0)}_{\mu \nu} +
\epsilon \vec{f}^{(1)}_{\mu \nu},
\nonumber \\
\vec{f}^{(0)}_{\mu \nu} &=& - \vec{B}_{\mu} \times \vec{B}_{\nu},
\nonumber \\
\vec{f}^{(1)}_{\mu \nu} &=& \frac{1}{2}
\left[ \vec{B}_{\nu}(x+ a e_{\mu}) - \vec{B}_{\nu}(x-a e_{\mu}) \right]
- \left( \mu \leftrightarrow \nu \right).
\end{eqnarray}
In the following we keep $g$ fixed and expand action and observable in terms of
$\epsilon$.
\begin{eqnarray}
S &=& S^{(0)} + \epsilon S^{(1)} + O(\epsilon^{2}),
\nonumber \\
S^{(0)} &=& \frac{1}{4} \sum_{x, \mu, \nu} (\vec{f}^{(0)}_{\mu \nu}(x) )^{2},
\nonumber \\
S^{(1)} &=& \frac{1}{2} \sum_{x, \mu, \nu}
\vec{f}^{(0)}_{\mu \nu}(x) \cdot \vec{f}^{(1)}_{\mu \nu}(x).
\end{eqnarray}
In Ref. \cite{kn:Krog92} by expanding
the fractal Wilson loop in $\epsilon$
we have obtained to lowest order the following result,
\begin{eqnarray}
<F^{(0)}_{P}> &=&  \frac{1}{Z}  \left[
\int d \vec{B}_{1} \cdots d \vec{B}_{4} \;
\cos [ \sqrt{g}  \mid \vec{B}_{1} \mid  ]
\exp \left( - \frac{1}{4} \sum_{\mu , \nu} (\vec{B}_{\mu}
\times \vec{B}_{\nu})^{2} \right) \right]^{N}
\nonumber \\
& \times &  \left[
\int d \vec{B}_{1} \cdots d \vec{B}_{4} \;
\cos [ g \mid \vec{B}_{1} \times \vec{B}_{2} \mid  ]
\exp \left( - \frac{1}{4} \sum_{\mu , \nu} (\vec{B}_{\mu}
\times \vec{B}_{\nu})^{2} \right) \right]^{M}
\nonumber \\
 Z &=& \left[
\int d \vec{B}_{1} \cdots d \vec{B}_{4} \;
\exp \left( - \frac{1}{4} \sum_{\mu , \nu} (\vec{B}_{\mu}
\times \vec{B}_{\nu} )^{2} \right) \right]^{N+M},
\end{eqnarray}
where $N$ is the number of links in $\partial P_{fract}$
and $M$ is the number of plaquettes in $P_{fract}$.
Now going to next order in $\epsilon$ the calculation gives
\begin{equation}
<F_{P}> = <F^{(0)}_{P}> + \epsilon <F^{(1)}_{P}> + O(\epsilon^{2}),
\end{equation}
where
\begin{eqnarray}
<F^{(1)}_{P}> &=&
\frac{1}{Z}
\int d \vec{B}_{1} \cdots d \vec{B}_{4} \;
\mid \vec{B}_{1} \mid
\sin [ \sqrt{g} \mid \vec{B}_{1} \mid /2]
\exp \left( - \frac{1}{4} \sum_{\mu , \nu} (\vec{B}_{\mu}
\times \vec{B}_{\nu})^{2} \right)
\nonumber \\
& \times &
\int d \vec{B}_{1} \cdots d \vec{B}_{4} \frac{N}{4}
\left[ \frac{1}{\sqrt{g}} \mid \vec{B}_{1} \times \vec{B}_{2} \mid
\sin [ g \mid \vec{B}_{1} \times \vec{B}_{2} \mid  /2]
+ \sqrt{g} \cos [ g \mid \vec{B}_{1} \times \vec{B}_{2} \mid  /2 ] \right]
\nonumber \\
& \times &
\exp \left( - \frac{1}{4} \sum_{\mu , \nu} (\vec{B}_{\mu}
\times \vec{B}_{\nu})^{2} \right)
\nonumber \\
& \times &
\left[ \int d \vec{B}_{1} \cdots d \vec{B}_{4} \;
\cos [ \sqrt{g} \mid  \vec{B}_{1} \mid ]
\exp \left( - \frac{1}{4} \sum_{\mu , \nu} (\vec{B}_{\mu}
\times \vec{B}_{\nu})^{2} \right) \right]^{N-1}
\nonumber \\
& \times &  \left[
\int d \vec{B}_{1} \cdots d \vec{B}_{4} \;
\cos [ g \mid  \vec{B}_{1} \times \vec{B}_{2} \mid ]
\exp \left( - \frac{1}{4} \sum_{\mu , \nu} (\vec{B}_{\mu}
\times \vec{B}_{\nu})^{2} \right) \right]^{M-1}
\end{eqnarray}
This shows that $<F_{P}>$ up to first order in $\epsilon$ behaves as
\begin{equation}
<F_{P}> = const \times \exp[ -\lambda M -\delta N]
\longrightarrow_{a/L \rightarrow 0}
const \times
\exp[ -(\lambda + \delta) \frac{2}{3} \frac{A_{P}}{a^{2}} ],
\end{equation}
i.e., an area law behavior.

\section {Gauge invariance}
Now we want to show that $<F_{P}>$ is gauge invariant also to first order
in $\epsilon$.
In terms of the rescaled variables, under a local gauge transformation
$G(x) = \exp[ i \vec{\chi}(x) \cdot \vec{\tau} ]$, the gauge field transforms
according to
\begin{equation}
B_{\mu}(x) \rightarrow B'_{\mu}(x) = G(x) B_{\mu}(x) G^{-1}(x)
- i \epsilon G(x) \left[ G^{-1}(x+ae_{\mu}) - G^{-1}(x-ae_{\mu}) \right].
\end{equation}
This can be expressed as
\begin{equation}
\vec{B}_{\mu}(x) \rightarrow
\vec{B}'_{\mu}(x) = O(x) \vec{B}_{\mu}(x) - \epsilon \; \vec{R}_{\mu}(x),
\end{equation}
where $O$ is a rotation  matrix and $\vec{R}$ is a translation
vector in color space.
Both, $O$ and $\vec{R}$, depend on $G(x)$.
Its explicit dependence can be computed, but it is not relevant for the
following considerations.
The fractal Wilson loop has been computed up to first order in $\epsilon$
in eqs.(7,8).
Analyzing the individual terms, one finds that
they are given by a functional integral
with the measure $\left[ \prod dB^{a}_{\mu}(x) \right]$
such that the integrands depend only on the modulus of the fields
(it does not contain, e.g., products of fields corresponding to different
lattice sites).
The integration domain can be chosen spherically symmetric
$\mid \vec{B}_{\mu}(x) \mid \leq M$.
Consequently, the functional integral is invariant under any rotation.
Thus up to first order, $<F_{P}>$ is invariant under rotations, i.e., the first
part of the gauge transformation.

Now let us consider the second part, i.e., the translation.
Evidently, $\epsilon <F_{P}>^{(1)} $ is already of first order, thus
a translation of the gauge fields by $\epsilon \; \vec{R}_{\mu}$
does not change anything to first order.
So finally let us look at $<F_{P}^{(0)}>$ under the translation.
Let us consider a one-dimensional integral
\begin{equation}
<f> = \int_{-m}^{+m} dx f(x),
\end{equation}
and a translation $f(x) \stackrel{T}{\rightarrow} f(x+\delta)$,
\begin{equation}
<f^{T}> = \int_{-m}^{+m} dx f(x+\delta) = <f> + \delta f(m) - \delta f(-m)
+ O(\delta^{2}).
\end{equation}
Under the assumption that $f$ is symmetric at the boundaries,
$f(m)=f(-m)$, one has $<f^{T}>=<f> + O(\delta^{2})$, i.e., invariance to first
order. The integrands of the functional integrals $F^{(0)}, F^{(1)}$
are rotationally invariant, hence also symmetric at the boundaries.
This establishes that $<F_{P}^{(0)}>$ is also invariant under the translations
given by eqs.(12,13).
Hence in summary, the fractal Wilson loop $<F_{P}>$ is invariant under local
gauge transformations also in first order of the expansion parameter
$\epsilon$.

\bigskip

{\bf Acknowledgement}
H.K. is grateful to J. Polonyi, H. Markum, I. Stamatescu and
V. Branchina for helpful discussions.
H.K. was supported by NSERC Canada and FCAR Qu\'ebec.

\newpage
\begin{flushleft}
{\bf Figure Caption}
\end{flushleft}
\begin{description}
\item[{Fig.1}]
Construction of fractal Wilson loop. L/a=8.
\end{description}
\end{document}